\documentclass[prl, twocolumn, showpacs, amsmath, amssymb, superscriptaddress]{revtex4}
\usepackage{graphicx}
\usepackage{graphics}
\usepackage{dcolumn}
\usepackage{bm}
\usepackage{amssymb}
\include{commands.tex}

\newcommand {\ket} [1] {| #1 \rangle}
\newcommand {\bkt} [1] {\langle #1 \rangle}

\newcommand {\pd} [2] {\frac{\partial #1}{\partial #2}}
\newcommand {\td} [2] {\frac{d #1}{d #2}}

 \newcommand {\beq}{\begin{equation}}
\newcommand {\eeq}{\end{equation}}
\newcommand {\bea}{\begin{eqnarray}}
\newcommand {\eea}{\end{eqnarray}}
\begin{document}
\title{Semiclassical spin transport in spin-orbit coupled bands}
\author{Dimitrie Culcer}
\affiliation{Department of Physics, The University of Texas at Austin, Austin TX 78712-1081}
\affiliation{International Center for Quantum Structures, Chinese Academy of Sciences, Beijing 100080, China}
\author{Jairo Sinova}
\affiliation{Department of Physics, Texas A\&M University, College Station, TX 77843-4242}
\author{N. A. Sinitsyn}
\affiliation{Department of Physics, Texas A\&M University, College Station, TX 77843-4242}
\author{T. Jungwirth}
\affiliation{Institute of Physics ASCR, Cukrovarnick\'a 10, 162 53
Praha 6, Czech Republic } \affiliation{School of Physics and
Astronomy, University of Nottingham, Nottingham NG7 2RD, United
Kingdom}
\author{A. H. MacDonald}
\affiliation{Department of Physics, The University of Texas at Austin, Austin TX 78712-1081}
\author{Q. Niu}
\affiliation{Department of Physics, The University of Texas at Austin, Austin TX 78712-1081}
\affiliation{International Center for Quantum Structures, Chinese Academy of Sciences, Beijing 100080, China}
\date{\today}
\begin{abstract}
Motivated by recent interest in novel spintronics effects, we develop a semiclassical theory of spin transport that is valid for spin-orbit coupled bands. Aside from the obvious convective term in which the average spin is transported at the wavepacket group velocity, the spin current has additional contributions from the wavepacket's spin and torque dipole moments. Electric field corrections to the group velocity and carrier spin contribute to the convective term. Summing all terms we obtain an expression for the intrinsic spin-Hall conductivity of a hole-doped semiconductor, which agrees with the Kubo formula prediction for the same quantity. We discuss the calculation of spin accumulation, which illustrates the importance of the torque dipole near the boundary of the system.
\end{abstract}
\maketitle
Electrical control of spins in systems with spin-orbit interactions is of basic interest and has great potential in semiconductor spintronics \cite{Prinz,Wolf,Das}. In recent years, steady progress has been made towards realization of convenient semiconducting ferromagnets and spin injection into semiconductors from ferromagnetic metals \cite{Schmidt,OhnoI,Fiederling,OhnoII,Parkin,Jonker}. The spin transport theory presented in this Letter is motivated generally by current interest in novel spin-related transport effects in semiconductors, and particularly by interest in various schemes that generate spin-polarized currents \cite{Nagaosa,sinova,spincurrent,Dyakonov,Hirsch,Zhang, Halperin}. Using a semiclassical wavepacket approach, we find that the spin current can be expressed as the sum of several physically transparent terms which are grouped together in a Kubo formula description. As an example, we use our theory to derive an expression for the intrinsic spin-Hall conductivity \cite{Nagaosa,sinova} of a hole-doped semiconductor.

Semiclassical formulations of transport theory exploit the smooth variation of transport fields on atomic length scales. Previous semiclassical theories of {\em spin} transport \cite{Dyakonov,QZh,Dugaev,Lopes,Fabian} have not accounted explicitly for intrinsic spin-orbit interaction in the crystal apart from, occasionally, its role in the relaxation of non-equilibrium spin-polarizations. In this Letter, we apply the wavepacket approach introduced by Sundaram and Niu \cite{Sundaram}, which captures the consequences of the wavevector dependence of the Bloch spinors, to treat spin transport in spin-orbit coupled bands.
This wavepacket approach has already been successful in describing the anomalous Hall effect in ferromagnetic semiconductors \cite{Tomas} and transition metals \cite{Yao}, interpreting it as a consequence of the Berry-phase correction to the group velocity induced by the intrinsic spin-orbit interaction.
We show here that the Hall spin current in response to an electric field is non-zero even in paramagnetic systems and that, in addition to the Berry phase term evaluated in a recent paper \cite{Nagaosa}, other contributions must be taken into account. First of all, there is a contribution from the electric field correction to the average spin orientation of a wavepacket. In addition, there are also contributions from the spin dipole and torque dipole of a carrier, which arise from the fact that spin and torque distribution within a wavepacket generally differ from that of the charge. Including all these contributions, we obtain a total semiclassical spin current which is in agreement with the Kubo formula expression for the same quantity. We show that nonequilibrium spin polarization near the sample edge is driven not by the spin current alone but by the sum of the spin current and torque dipole density.

The semiclassical dynamics of each spin-charge carrier in a non-degenerate band is described by a wavepacket, whose charge centroid has coordinates (${\bf r}_c, {\bf k}_c$). Wavepacket construction is thoroughly explained in \cite{Sundaram,AM}. When expanded in the basis of Bloch eigenstates, the wavepacket has the form:
\beq
|w\rangle=\int d^3k \; a({\bf k}, t) e^{i{\bf k}\cdot\hat{\bf r}}|u({\bf r}_c, {\bf k}, t)\rangle .
\eeq
In the above, the wavefunctions $\ket{u}$ contain \cite{Culcer} corrections linear in the electric field. They form a complete set and retain the Bloch periodicity. The function $a({\bf k}, t)$ is a narrow distribution sharply peaked at ${\bf k}_c$, and its phase specifies the center of charge position ${\bf r}_c$. The size of the wavepacket in $k$-space must be considerably smaller than that of the Brillouin zone. In real space, this implies that the wavepacket must stretch over many unit cells.

In the presence of a uniform electric field, the semiclassical equations of motion for a non-degenerate band read \cite{Sundaram}:
\begin{eqnarray}
\hbar \dot {\bf r}_c = \pd{\varepsilon}{{\bf k}_c} - e{\bf E} \times {\bf \Omega}
\,\,\,{\rm and}\,\,\, \hbar \dot {\bf k}_c
=e{\bf E},
\end{eqnarray}
where $e$ is the carrier charge, $\varepsilon$ is the band dispersion, and ${\bf \Omega}$ is the Berry curvature of the Bloch state \cite{Sundaram}. Henceforth, ${\bf k}_c$ will be abbreviated to {\bf k}. The effect of the electric field is thus twofold: it drives the center of the wavepacket in $k$-space, and it gives rise to a non-adiabatic correction to the wavefunctions, which mixes the bands at each {\bf k}.

Following the strategy of Boltzmann transport theory, we consider a collection of particles described by a phase space distribution $f({\bf r}_c, {\bf k}, t)$. This distribution can drift according to the semiclassical equations of motion (2), and can also change due to collisions:
\beq
\label{Boltz}
\pd{f}{t} + \dot{\bf r}_c\cdot\pd{f}{{\bf r}_c} + \dot{\bf k}\cdot\pd{f}{{\bf k}} = \td{f}{t}|_{coll}.
\eeq
The collision term on the right hand side may be modeled by relaxation times or more accurately by collision integrals as usual.

The spin density distribution is defined as
\beq
S({\bf r}, t) \equiv \int\!\int d^3k d^3r_c f({\bf r}_c, {\bf k}, t)\bkt{\delta({\bf r} - {\bf \hat r})\hat s},
\eeq
where $\hat s$ is an arbitrary component of the spin operator, and the bracket indicates quantum mechanical average over the wavepacket with charge centroid $({\bf r}_c, {\bf k})$. Further analysis of this distribution will be facilitated by making the analogy with the standard coarse graining of electrodynamics in material media \cite{Jackson}. Our wavepackets play the role of `molecules', whose size will be taken as much smaller than the length scale of the distribution function. We are thus allowed to view the $\delta$-function in the above definition of the spin density as a sampling function with a width somewhere between the microscopic scale of the wavepackets and the macroscopic scale of the distribution function. We can therefore write it as $\delta[({\bf r} - {\bf r}_c) - ({\bf \hat r} - {\bf r}_c)]$ and expand it around ${\bf r}_c$, keeping only the first order term. Performing the integral over ${\bf r}_c$, the spin density can be re-expressed in the following form:
\beq
S = \int d^3k f\bkt{\hat s} - \nabla\cdot\int d^3k f {\bf p}^s
\eeq
where $f = f({\bf r}, {\bf k}, t)$, and ${\bf p}^s= \bkt{(\hat{\bf r} - {\bf r}_c)\hat s}|_{{\bf r}_c = {\bf r}}$ is the spin dipole. The two terms can be regarded as monopole and dipole contributions. The second term is analogous to the contribution to the charge density in electrodynamics from the divergence of the polarization.

Spin is in general not conserved, and for what follows it will be useful to define a quantity, which we shall call the torque density, in order to include the rate of change of spin into our discussion of transport:
\beq
{\cal T}({\bf r}, t) \equiv \int\!\int d^3k d^3r_c f({\bf r}_c,
{\bf k}, t)\bkt{\delta({\bf r} - {\bf \hat r}){\hat \tau}},
\eeq
in which ${\hat \tau}$ is understood as $\frac{i}{\hbar}[\hat H, \hat s]$ and $\hat H$ is the Hamiltonian. Following the steps outlined above, the torque density becomes:
\beq
{\cal T} = \int d^3k f\bkt{\hat \tau} - \nabla\cdot\int d^3k f {\bf p}^\tau,
\eeq
with the torque dipole ${\bf p}^\tau = \bkt{({\bf \hat r} - {\bf r}_c){\hat \tau}}|_{{\bf r}_c = {\bf r}}$.

We evaluate the spin-current using the microscopic spin-current operator and the semiclassical distribution function:
\beq
{\bf J}^s({\bf r}, t) \equiv \int\!\int d^3k d^3r_c \; f({\bf r}_c, {\bf k}, t)\bkt{\delta({\bf r} - {\bf \hat r}){\bf \dot {\hat r}}{\hat s}}.
\eeq
Throughout this paper, symmetrization of products of noncommuting operators is implied. After expanding, the spin current takes the form:
\beq
{\bf J}^s = \int d^3k f\bkt{{\bf \dot{\hat r}}{\hat s}} - \nabla\cdot\int d^3k f\bkt{({\bf {\hat r}} - {\bf r}){\bf \dot{\hat r}} {\hat s}}
\eeq
For a homogeneous system, where the distribution function is independent of position, the gradient term vanishes, and it is permissible to use Bloch states
(which may be regarded as the limit of very wide wavepackets) to evaluate the carrier spin current $\bkt{{\bf \dot{\hat r}}{\hat s}}$.  Since the Bloch states contain first order correction in the field, this can in general yield an overall linear-in-field spin current even with the equilibrium distribution function.  This intrinsic spin current has been evaluated for a number of systems recently, and identical results are obtained with the semiclassical approach developed here.

To illuminate the underlying physics, we decompose the carrier spin current into a number of terms:
\beq
{\bf J}^{s} = \int d^3k f ({\bf \dot r}_c\bkt{\hat s}+ \td{{\bf p}^s}{t} - {\bf p}^\tau).
\label{js}
\eeq
The first contribution is convective, arising from the fact that the total spin is transported as the wavepacket moves. The second comes from the rate of change of the spin dipole, while the third is from the torque dipole.  This decomposition makes it possible to compare the Kubo formula result with those based on various heuristic arguments. The authors of \cite{Nagaosa} restricted their scope to the convective term and considered only the Berry phase contribution to the charge center velocity $\dot {\bf r}_c$. The present semiclassical decomposition allows us to recognize the missing terms due to the spin and torque dipoles, as well as a field correction to the carrier spin in the convective term. The approach of \cite{Nagaosa} would give a zero result for the Rashba model, whereas the Kubo formula approach of \cite{sinova}, which agrees with (\ref{js}), yields a nonzero spin Hall current for this model. Interestingly, the spin Hall current in the Rashba model can be obtained exactly from a heuristic argument based on a velocity and field dependent correction to the carrier spin as discussed in \cite{sinova}. This approach is, however, applicable only to single-band models with wavevector-dependent Zeeman coupling.

The spin density and current satisfy the following equation of continuity:
\beq
\label{continuity}
\pd{S}{t} + \nabla\cdot{\bf J}^{s} = {\cal T} + \int d^3k\td{f}{t} \bkt{\hat s}.
\eeq
It is seen that the torque density appears in the source, accounting for the spin non-conserving terms in the Hamiltonian, and acting as a bulk mechanism for spin generation. The second term accounts for the effect of collisions.

The source can be decomposed into intrinsic and extrinsic contributions, depending on the equilibrium and non-equilibrium parts of the distribution respectively. If we restrict our attention to homogeneous sytems the torque density is simply $f\bkt{\hat \tau}$. We find that this term is always first order in the electric field, and is given by $f{e\over \hbar}{\bf E}\cdot\pd{\bkt{\hat s}}{{\bf k}}$. We are thus justified in replacing $f$ by its equilibrium value $f_0$, in which case this term is purely intrinsic. The second term in the source, which depends on the nonequilibrium shift in the distribution function, is entirely extrinsic.

Our formalism so far applies to independent non-degenerate bands, and for the Rashba model its predictions are in agreement with \cite{sinova}. There exists a parallel formalism for coupled degenerate bands, which yields the same results as given above \cite{Culcer}. In this case, the distribution function becomes a density matrix, while $\bkt{\hat s}$, $\bkt{\hat \tau}$, $\dot{\bf r}_c$, ${\bf p}^s$ and ${\bf p}^\tau$ are replaced by matrices. To find the macroscopic expectation values, one traces over the density matrix. This formalism can be applied, for example, to the spherical four-band Luttinger Hamiltonian,
\beq
H_0 = \frac{\hbar^2}{2m}[(\gamma_1 + \frac{5}{2}\gamma_2)k^2 -
2\gamma_2({\bf k}\cdot \hat {\bf J})^2] ,
\eeq
where $\hat {\bf J}$ is the operator for angular momentum 3/2 and $\gamma_1$, $\gamma_2$ are the Luttinger parameters. The Bloch states are eigenstates of the angular momentum projection in the {\bf k} direction, $\hat J_k$. The four bands are split (for finite $k$) into two degenerate manifolds with $\hat J_k=\pm 3/2$ (heavy holes) and $\hat J_k=\pm 1/2$ (light holes).

Let us take a closer look at the source term, using the four-band model as an illustration. This discussion applies to either of the heavy and light-hole manifolds. In equilibrium, the density matrix is diagonal and equal to $f_0$ for each band. The mean spin in the $z$-direction is $\bkt{\hat s_z}_{\pm 3/2} = \pm\frac{\hbar k_z}{2k}$ for the heavy holes, and it is $\bkt{\hat s_z}_{\pm 1/2} = \pm\frac{\hbar k_z}{6k}$ for the light holes. The spin expectation values have opposite signs in the two bands, so that, when averaged over the equilibrium density matrix, the intrinsic term in the source, $f\bkt{\hat \tau}$, will vanish. The intrinsic source ${\cal T}$ therefore vanishes in the bulk for this system.

In the relaxation time approximation, the collision term in (\ref{Boltz}) is given by \cite{note}:
\beq
\label{coll}
\td{f}{t}|_{coll} = \frac{f_0 - \frac{1}{2}Tr f}{\tau_p}I - \frac{Tr(f{\bf \sigma})\cdot{\bf \sigma}}{\tau_s},
\eeq
where $\tau_p$ and $\tau_s$ are the momentum and spin relaxation times respectively, $I$ is the identity matrix and ${\bf \sigma}$ is the vector of Pauli spin matrices. In the extrinsic term in the source, the part depending on the momentum relaxation time will also cancel between the two bands, leaving us with just the contribution coming from the second term on the right hand side of (\ref{coll}). The equation of continuity is then:
\beq
\label{cont}
\pd{S}{t} + \nabla\cdot{\bf J}^{s} = \frac{ - S}{\tau_s} - \nabla\cdot{\bf P}^\tau,
\eeq
where ${\bf P}^\tau = \int d^3k f {\bf p}^\tau$ is the torque dipole density. The two divergences will vanish in the bulk if the sample is homogeneous.

We will now take a closer look at the spin current, making further use of the four-band model for the spin-orbit coupled valence bands of a zincblende semiconductor. In previous work \cite{Dyakonov,Hirsch,Zhang}, extensive discussions have been devoted to the extrinsic part of the spin current, which is given by the zero-field carrier velocity and spin integrated over the non-equilibrium part of the distribution. Here we will concentrate on the intrinsic part of the spin current, coming from the field correction to the carrier spin current integrated over the equilibrium distribution. In order for this term to be dominant, scattering must be strong enough to keep the distribution function close to its equilibrium value, and weak enough to limit inter-band mixing. This is therefore opposite to the limit of Dyakonov-Perel \cite{DyakonovJETP} relaxation of spin in the weakly spin-orbit split bands of crystals.

The two-fold degeneracy of both the heavy and light hole manifolds implies that, in the presence of an electric field, however weak it may be, mixing within the degenerate manifold will occur in general. Fortunately, for the heavy holes the $\hat J_k=\pm 3/2$ bands do not mix to first order in the electric field, and we can apply the nondegenerate band formalism. The $s_z$ dipole moment for the heavy holes is found to be ${\bf p}^{s_z}_{h} = - \frac{\hbar{\bf k}\times\hat{\bf z}}{4k^2}$. The torque dipole is, after an angular average, ${\bf p}^{\tau_z}_{h} = -\frac{e{\bf E}\times\hat{\bf z}}{6k^2}$. The spin and torque dipoles are equal for both heavy hole bands. For the convective part of carrier spin current, in addition to a field correction to the carrier velocity due to the Berry phase \cite{Nagaosa}, we obtain a term which is due to the change in the spin expectation value induced by the electric field and has the form $\frac{1}{\hbar}\pd{\varepsilon}{{\bf k}}{\bf E}\cdot\pd{\bkt{\hat s}}{{\bf E}}$. Using these results, we find the current for spin-$z$ component of a heavy-hole carrier to be
\beq
\label{jsh}
{\bf j}^{s}_{h} =(\frac{1}{4k^2} - \frac{\hbar^2}{6m_h\Delta} - \frac{1}{12k^2}  + \frac{1}{6k^2})e{\bf E}\times{\bf \hat z},
\eeq
where $\Delta = \epsilon_h - \epsilon_l$ is the energy difference between the heavy and light holes. The first two terms come from the convective part due to field corrections to the carrier velocity and spin respectively. The third term comes from the rate of change of the spin dipole, while the last one comes from the torque dipole. The heavy-hole carrier spin current can be simplified to:
\beq
{\bf j}^{s}_{h} =\frac{e{\bf E}\times{\bf \hat z}}{3 k^2}\frac{1}{1 - \frac{m_l}{m_h}}.
\eeq
For the light holes, we must consider field induced mixing between the two degenerate bands.  The details of this calculation will be deferred to a future publication \cite{Culcer}, we only quote the final result here which is very simple.  The spin current per carrier in the light hole manifold has the same form as for the heavy holes, and differs only by a minus sign. Integrating over $k$ and summing the contributions from all four bands, we arrive finally at the following expression for the total spin current: ${\bf J}^{s} = \sigma_{SH}{\bf E}\times{\bf \hat z}$, where the spin-Hall conductivity is given by:
\beq
\sigma_{SH} =
\frac{e}{3\pi^2}\frac{k_h-k_l}{1 - \frac{m_l}{m_h}}=
\frac{e}{3\pi^2}\frac{k_h}{1 + \sqrt{\frac{m_l}{m_h}}}
\eeq
where $k_h$ and $k_l$ are the Fermi wavevectors for the heavy and light holes respectively. Separate calculations based on the Kubo formula by the present authors and by Murakami {\it et al.} \cite{holonomy} yield the same results.

Finally, we comment on the relationship between bulk spin currents and spin accumulation near the edge of the sample. A theory of spin accumulation must start from the spin density continuity equation (\ref{cont}). If the torque dipole density ${\bf P}^\tau$ were absent from this equation, then in the steady state the spin accumulation would be due only to the spin current. The presence of the torque dipole density modifies the expression for the spin accumulation, giving that
\beq
\int S dx = \tau_s (J^s_x + P^\tau_x).
\eeq
We have already discussed the response of the spin current to an electric field, and after a similar calculation for the torque density we find that
\beq
J^s_x + P^\tau_x = \frac{eE_y}{3\pi^2}\frac{k_h}{1 + \sqrt{\frac{m_l}{m_h}}}(\frac{1}{2} - \frac{m_l}{2m_h} - \sqrt{\frac{m_l}{m_h}}).
\eeq
Using $n = 2.4\times 10^{11}$ cm$^{-2}$ and an electric field of 20000V/cm as typical values, the spin current is $-10^{25}$ spins per unit area per second. We take the spin relaxation time to be $\tau_s = 30$ps \cite{jpreltime} and a unit cell size of $6.3 \AA$, and we obtain a spin accumulation of $1.2\times 10^{-4}$ spins per {\it unit cell area}. This is a measurable effect as discussed in \cite{Nagaosa,sinova}.

The authors gratefully acknowledge stimulating discussions and communications with S. Murakami, N. Nagaosa, and S-C. Zhang. This work was supported by the DOE through grant DE-FG03-02ER45958, by the Welch Foundation, and by the Grant Agency of the Czech Republic under grant 202/02/0912. NAS was supported by NSF under grant DMR0072115.


\begin{references}
\bibitem{note}We have defined the collision term in analogy with Eq. (2) in [16]. However, we have made some simplifications which are valid for degenerate bands.
\end{references}
\end{document}